\documentclass[runningheads]{llncs}
\usepackage{graphicx}
\usepackage{comment}

\usepackage{microtype}
\usepackage{graphicx}
\usepackage{subfigure}
\usepackage{booktabs} % for professional tables
\usepackage{balance}
\usepackage[figuresleft]{rotating}

\usepackage{xcolor}

\usepackage{graphicx}
\usepackage{amsmath}
\usepackage{amssymb}
\usepackage{booktabs}

\usepackage{xspace}
\graphicspath{{fig/}}

\usepackage{hyperref}

\newcommand{\webname}{\href{https://reproducedpapers.org/}{\tt ReproducedPapers.org}}

\begin{document}
%
%\title{\webname\thanks{Actual name of website is hidden for anonymous peer review.}: Openly teaching and structuring  machine learning reproducibility}

\title{\webname: Openly teaching and structuring  machine learning reproducibility}
\titlerunning{Openly  teaching and structuring  ML reproducibility}

\author{Burak Yildiz\orcidID{0000-0001-9932-4221} \and
Hayley Hung \and
Jesse H. Krijthe\orcidID{0000-0003-3435-6358} \and
Cynthia C. S. Liem\orcidID{0000-0002-5385-7695} \and
Marco Loog\orcidID{0000-0002-1298-8461} \and
Gosia Migut \and
Frans Oliehoek \and
Annibale Panichella\orcidID{0000-0002-7395-3588} \and
Przemysław Pawełczak\orcidID{0000-0002-1302-1148} \and
Stjepan Picek\orcidID{0000-0001-7509-4337} \and
Mathijs de Weerdt\orcidID{0000-0002-0470-6241} \and
Jan van Gemert\orcidID{0000-0002-3913-2786}
}
\authorrunning{B. Yildiz et al.}
\institute{Delft University of Technology, Postbus 5, 2600 AA Delft, The Netherlands}
\maketitle              % typeset the header of the contribution

\begin{abstract}

We present \webname: an open online repository for teaching and structuring machine learning reproducibility. We evaluate doing a reproduction project among students and the added value of an online reproduction repository among AI researchers. We use  anonymous self-assessment surveys and obtained 144 responses. Results suggest that students who do a reproduction project place more value on scientific reproductions and become more critical thinkers.  Students and AI researchers agree that our online reproduction repository is valuable.

%\jvg{ We submit here \url{https://participatoryml.github.io/} }


\end{abstract}
\section{Introduction}
\label{intro}

% Move to later: In line with the National Information Standards Organization (NISO) and ACM, we define reproducibility to be not just the replication of existing results from a author's original research code but a full reimplementation according the written description of the paper. 

Reproducibility is a cornerstone of science: if an experiment is not reproducible, we should question its conclusions. Yet, machine learning papers are lacking reproductions~\cite{Drummond,hutson2018artificial}.
Possible reasons may include a misaligned incentive between reproducing results and the short-term measures of career success associated with more `wins'~\cite{sculley2018winnersDurse} and publishing `novel' work~\cite{Lipton:2019:RPT:3336127.3316774}. Nevertheless, high-impact can be achieved, for instance, when a reproduction fails spectacularly, e.g.~\cite{dacrema_et_al_recsys_19,engstrom2020implementation,gorman2019standardSplits,henderson2018deepRLthatMatters,lin2019neuralHypeWeakBaselines,lucic2018gansEqual,melis2018onNLM,musgrave2020metricLearnRealCheck,riquelme2018deepBayesianShowdown}.
% For reproductions that do not fail, there is the excellent peer-reviewed journal ReScience C~\cite{rougier2018rescience} which upholds rigorous standards for reproductions or badging schemes by ACM~\cite{artifacts:online:acm:2020,boisvert:reproducibility:cacm:2019} and others~\cite{peng2011reproducible} which encourage authors to share data, code and have reproducible results. Such
Yet, these take colossal amounts of manual effort~\cite{anand20blackMagic,bonneel2020code,fursin:artifact:asplos:2020,raff2019step} or massive resources~\cite{lucic2018gansEqual,recht2019imagenet2imagenet}.
There are venues for publishing reproductions~\cite{colom2018overview,colom2015ipol,rougier2018rescience}, which are typically peer-reviewed and thus uphold various selection standards to guarantee quality. We argue that this emphasis on quality is a hurdle for sharing light-weight reproductions. Important and useful examples of light-weight reproductions include partial results, small variants on the algorithm, hyperparameter sweeps, etc. Low-barrier options are indeed available in workshop challenges~\cite{kerautret2019reproducible,pineau2019iclr} organized at conferences such as ICPR, NeurIPS, ICLR, or ICML. However, such avenues are hard to maintain on a long-term basis, as a workshop may or may not be organized. We argue that there is a need for a low-barrier and long-term venue for machine learning reproductions.% in the machine learning community.

%In addition to venues for sharing reproduction results, 
A complementary angle on low-barrier reproductions is to improve university student training. We should teach the next generation of machine learning practitioners the importance of the reproducibility of research work, as done in other computer science domains such as computer networking, where results reproduction is the means to learn new material~\cite{yan:reproducible:ccr:2017}. Doing a reproduction project in a course aligns with several important learning objectives for machine learning students. Among others, students (1) should be able to read, critique, and explain a scientific paper; (2) implement a method; (3) run, evaluate, investigate, and extend existing research or code; and (4) write clearly and concisely about code and methods. % Doing a reproduction project fits these objectives while letting
A reproduction project also lets students experience differences between published results and an implementation, which stimulates a critical attitude and allows reflections on the scientific process.  %In addition, it is an important first step towards developing ones own research questions that require students to have an intimate understanding of the nuances of an existing approach. Too often we see students take published work at face value and perhaps at a higher gold standard than is conducive to good critical thinking.

\begin{figure}
\centering
\includegraphics[width=\linewidth]{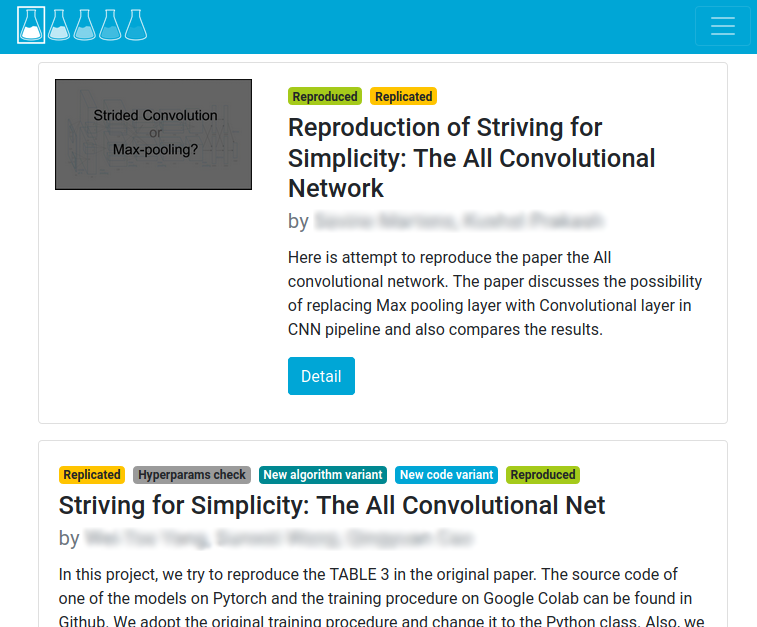} 
\caption{A screenshot of \webname. We allow multiple reproductions of the same original paper and investigations of several aspects, such as \emph{Reproduced}, \emph{Replicated}, \emph{Hyperparameter check}, etc. Our  online repository is user-centered: its sufficient if a user sees value in uploading some form of reproduction. Having such a repository is well-suited for students and adds structure to reproducibility in machine learning.}
%\pp{Suggestion: by cutting the last paper (Combinatorial Optimization...) from the figure we keep the information intact, while we add extra space to the paper}
\label{fig:screenshot}
\end{figure}

\newpage
In this paper, we align the benefits of an online reproduction repository with those of teaching reproducibility. We introduce \webname: an open, light-weight repository of reproduced papers which flexibly allows any sort of reproduction work, see 
Figure~\ref{fig:screenshot}. This repository benefits the research community while at the same time being well-equipped at accepting contributions from students. Although the standard of student reproductions might be lower than those required for peer reviewed reproductions, they can still give valuable insights such as clarifying which parts are difficult to implement or identifying the reproducibility level of elements. Such online reproductions are a low-threshold portfolio-building opportunity, which in turn may prove a valuable incentive to start doing more reproductions, as well as an opportunity to facilitate sharing reproductions that otherwise would not have been shared. 

Our online repository shares traits with other light-weight, bottom-up, grassroots community efforts such as \emph{ArXiv}~\cite{arxiv}, \emph{Open Review}~\cite{soergelICML2013openScholarship}, and \emph{Papers with Code}~\cite{paperswithcode}. Other efforts on facilitating reproducibility include software for reproducible and reusable experiments~\cite{paganini2020dagger}, open specification neural network diagrams~\cite{marshall2018diagrammatic},  and a framework for automatic parsing of deep learning research paper to generate the implementation~\cite{sethi2018dlpaper2code}. Similar to these approaches, in our work, we combine the traits from online repositories with those of tools facilitating reproducibility by providing an online repository that facilitates teaching as well as structuring reproducibility.

We make the following contributions. 1. We propose a new online reproduction repository; 2. We conduct a proof of concept with students from an MSc Deep Learning course to perform a reproduction project and populate the repository; 3. We evaluate the usefulness of the repository among  AI researchers and the learning objectives among students by anonymous surveys.

\section{The online reproduction repository}
\label{sec:website}

We performed a proof of concept experiment with a reproducibility project for students of the MSc Deep Learning course taught by this paper's last author at Delft University of Technology (TU Delft). We solicited relevant papers among university staff and ensured that (i) data is available, (ii) it is clear which table or figure to reproduce, and (iii) the computational demands are reasonable. Students were also allowed to themselves suggest a paper to reproduce. On their paper of choice, they worked in groups of 2 to 4, for 8 weeks, for approximately one-third of their studying time (i.e., about 13 hours a week). For grading, students submitted a blog in PDF and also the URL of an online version of their blog to \webname~to populate the repository. For students who do not wish to share a blog with the world, we offer a private option, which is only visible to course administrators. The option to publicly blog about reproducing machine learning provides an simple opportunity for students to build an online portfolio while simultaneously incentivizing making reproductions.

\begin{table}
\centering
\begin{tabular}{p{0.3\linewidth}p{0.7\linewidth}} \toprule
 Aspect & Description \\ \midrule
$\bullet$ \small{Replicated} & \small{A full implementation from scratch without using any pre-existing code.} \\
$\bullet$ \small{Reproduced} & \small{Existing code was evaluated.} \\
$\bullet$ \small{Hyperparams check} & \small{New evaluation of hyperparameter sensitivity.} \\
$\bullet$ \small{New data} & \small{Evaluating new datasets to obtain similar results.} \\
$\bullet$ \small{New algorithm variant} & \small{Evaluating a different variant.} \\
$\bullet$ \small{New code variant} & \small{Rewrote/ported existing code to be more efficient/readable.} \\
$\bullet$ \small{Ablation study} & \small{Additional ablation studies.} \\ \bottomrule
\end{tabular}
\caption{Different aspects of reproduction which are highlighted as badges (see Figure~\ref{fig:screenshot}). 
%\fao{the badges are too small to see in the figure}
}
\label{tab:badges}
\end{table}

\newpage

\begin{figure}[b]
\centering
\begin{tabular}{cc}
\includegraphics[width=0.54\linewidth]{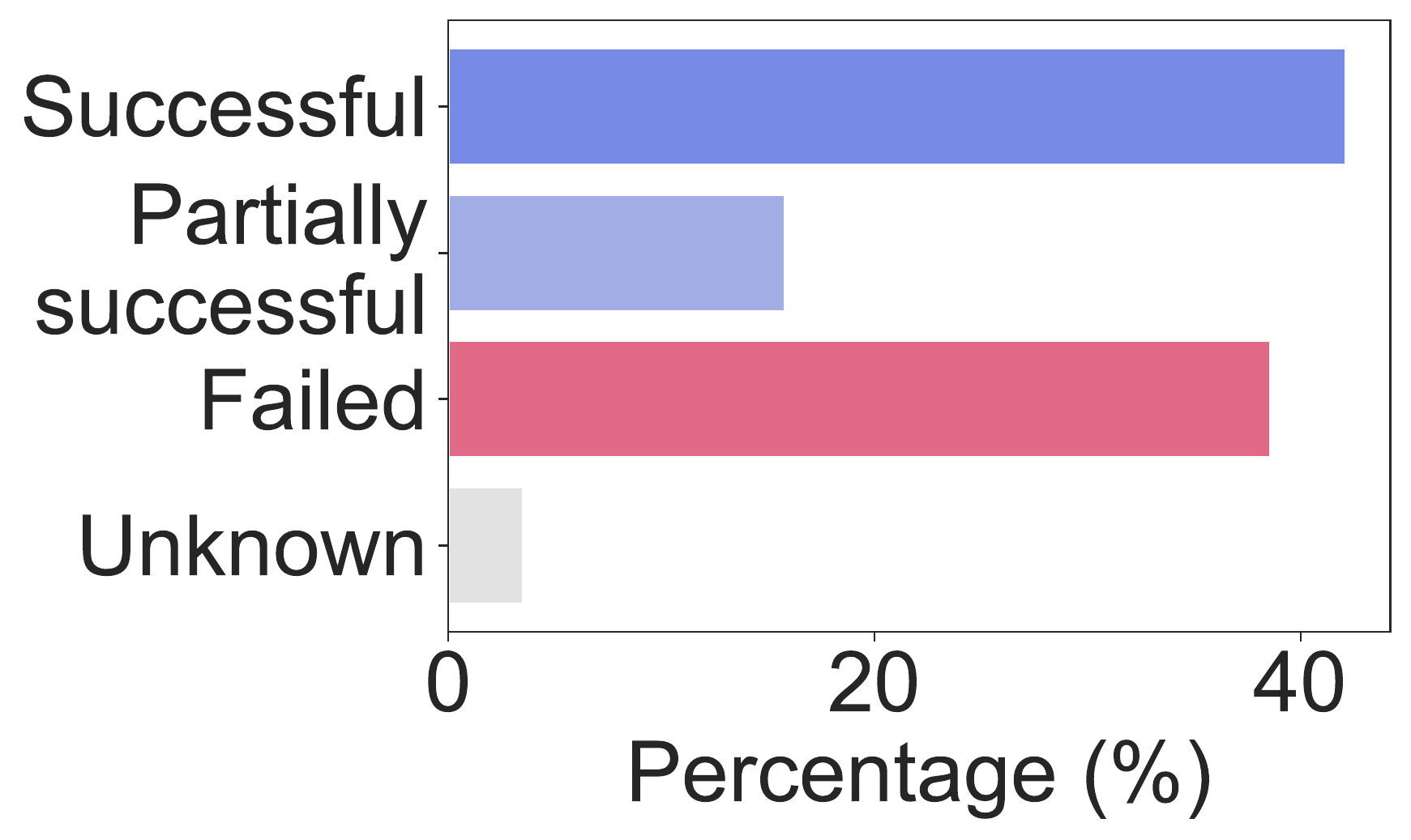} & \includegraphics[width=0.46\linewidth]{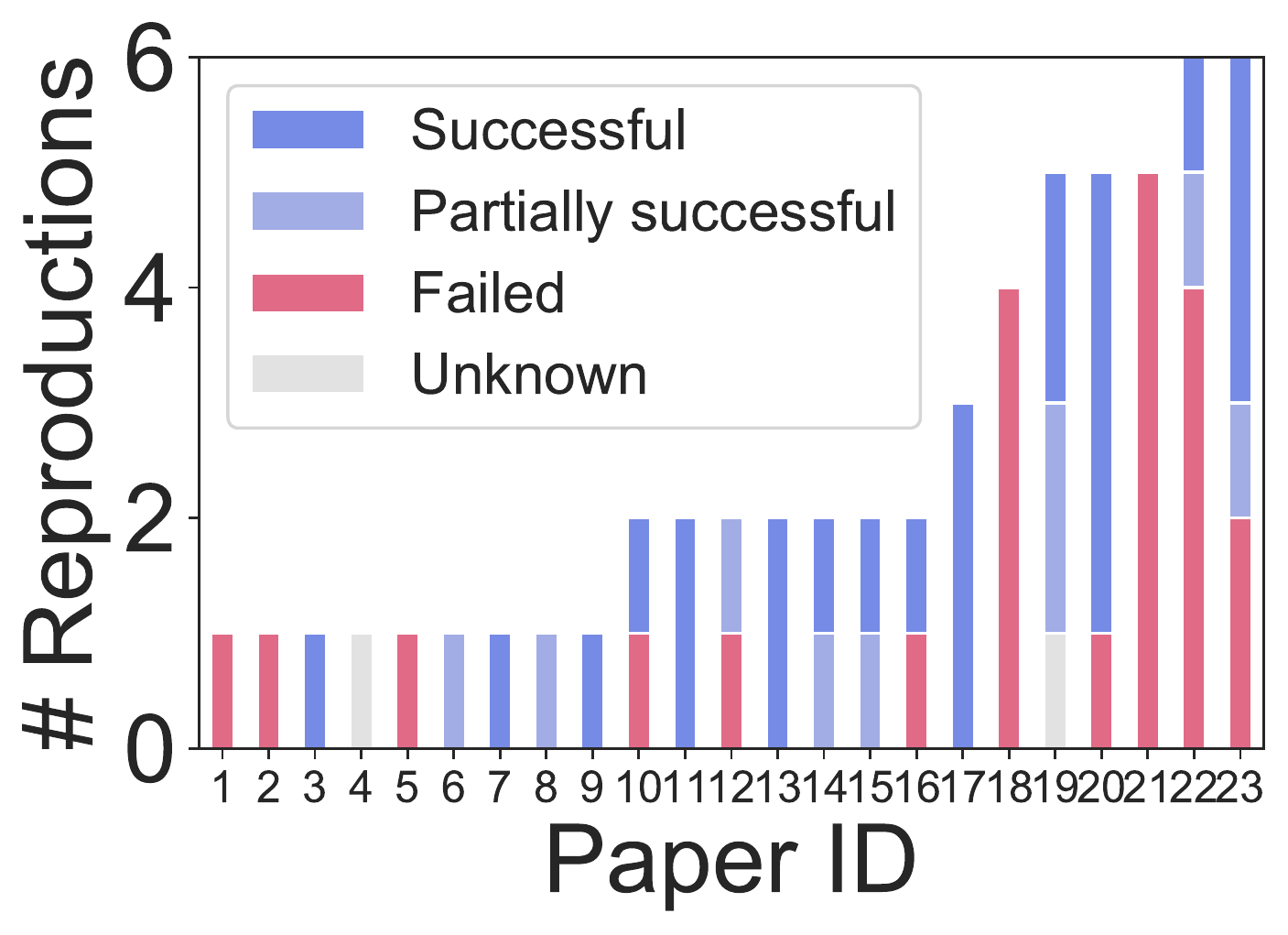} \\
(a) & (b)
\end{tabular}
\caption{Current \webname~statistics. (a) Reproduction success rates; (b) Number of reproductions per paper ID.}
%\pp{Suggestions to improve figures: bigger font for Fig b; Papers$\rightarrow$Paper Number}
\label{fig:successrates}
\end{figure}

%In addition to a heavy-weight re-implementation of a method from scratch, w
We explicitly allow for light-weight reproduction efforts such as evaluating existing code, checking only certain parts of the paper, proposing minor variations, doing hyperparameter sweeps, etc. Our current options (aspects) are shown in Table~\ref{tab:badges}, %To quickly categorize what type of reproduction is available we adopt the concept of `badges' from ACM~\cite{artifacts:online:acm:2020,boisvert:reproducibility:cacm:2019} and others~\cite{peng2011reproducible}. We propose a light-weight, bottom-up grassroots approach to reproduction and thus we do not aim to define a grand, overarching categorization of all possible reproduction options. 
and we will add others as the need arises.  Authors label their  reproduction with the relevant aspects themselves.

We developed \webname~in-house as a simple web application. It is implemented by this paper's first author, and its source code is available on GitHub\footnote{\url{https://github.com/CVLab-TUDelft/reproduced-papers}}.  Registering is necessary only when adding reproductions. Currently, the repository has 90 registered users and hosts 24 unique papers and 57 paper reproductions. Most papers have multiple reproductions, and only five reproductions are marked as private. The top-3 most-used aspects are \emph{Replicated} (32 times); \emph{Reproduced} (29 times) and \emph{Hyperparams check} (17 times). Figure~\ref{fig:successrates} whose data is derived from self-reported blog posts by users shows both success and failure rates to be around 40\%. % and some of the authors reported partial success and its rate is around 15\%. 

\begin{figure}[t]
\centering
\includegraphics[width=\linewidth]{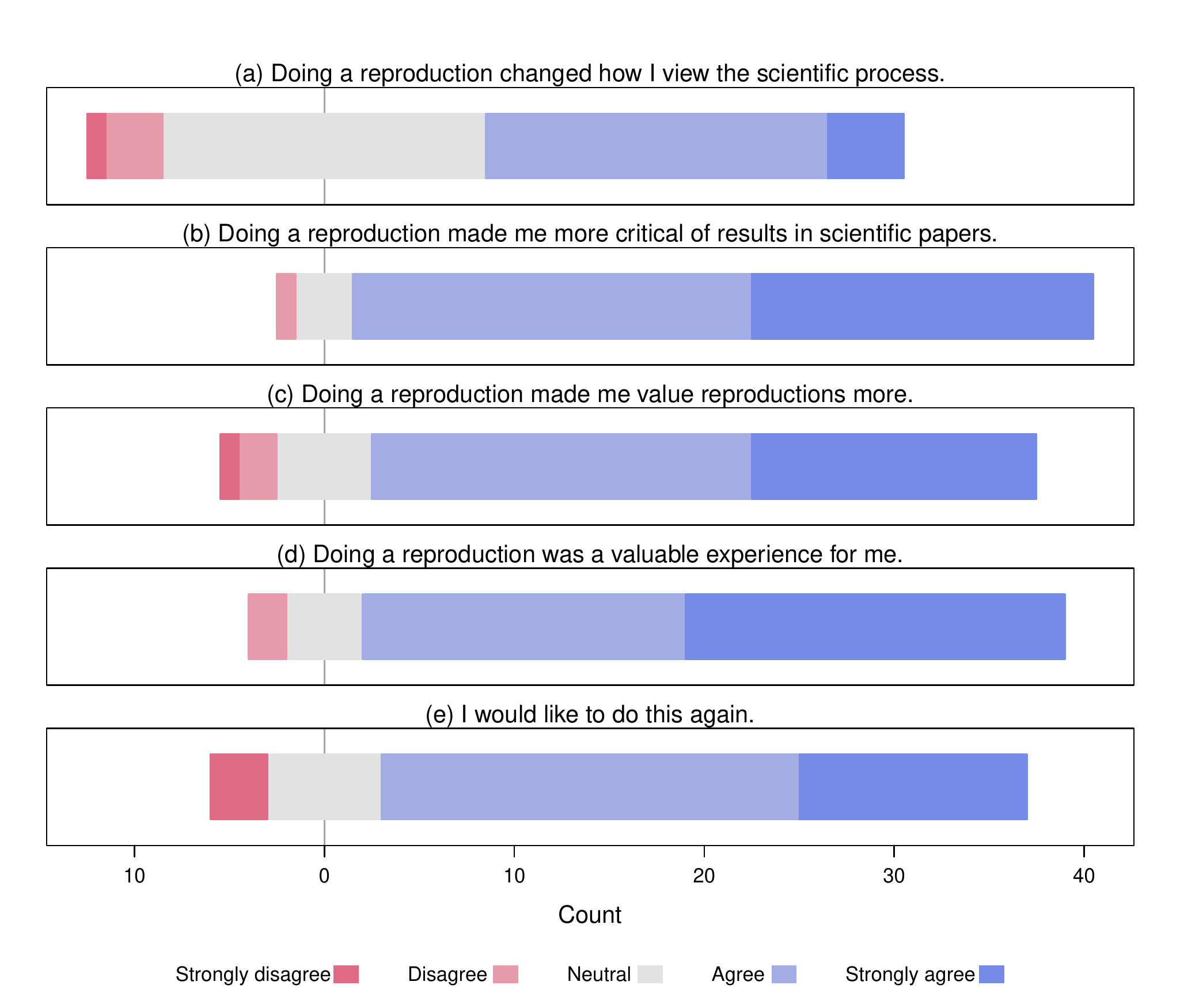} 
\caption{Responses to survey questions from students who contributed to \webname. Letting students themselves do a reproduction promotes a critical mindset (a and b), and teaches the value of scientific reproductions (c). In addition, the students considered it a positive experience (d,e). We conclude that these traits align with our learning objectives.}
\label{fig:studentresponses}
\end{figure}

\section{Survey analysis}

%\fao{"evaluate the educational aspect and usefulness" is a bit on the fuzzy side... Can we be more precise as to what we expect? Are there hypotheses? }
%\jvg{Tried to make them more explicit}

We evaluate student learning objectives and how AI researchers perceive our online reproduction repository by analyzing the results of small anonymous surveys for two groups: (i) students who recently added their reproduction  to our repository and (ii) anybody identifying her/himself working in AI. The second group was invited to the survey through social media and emails. Both groups share the same questions, where the students have five additional questions to evaluate education. The survey data is available at \webname\footnote{\url{https://reproducedpapers.org/survey-data.zip}}

We received a total of 144 responses: 43  from course students and 101 from third-party AI researchers all over the world. Of the latter, 87 identify themselves as a junior or senior researcher, and 14 as a student.

\subsection{Evaluating student learning objectives}

%\fao{2 objectives, but 3 figures? bit confusing. Perhaps add a 3rd objective?}
% \jvg{Yes, I agree, but I have 5 figs to fit, if I plot them in 1 row its too small :) )

The survey questions and results can be found in Figure~\ref{fig:studentresponses}. We evaluate the following objectives.

\textbf{Doing a reproduction project increases critical thinking.} Results in Figure~\ref{fig:studentresponses}(a) show that doing a reproduction taught most students something new about the scientific process. Figure~\ref{fig:studentresponses}(b)  suggests that students become more critical to published results.

\textbf{Doing a reproduction project makes students value reproductions more.} The results in Figure~\ref{fig:studentresponses}(c) indicate that after doing a reproduction, a great majority of students place more value on scientific reproductions.

\textbf{Students find a reproduction project a positive experience.}
The results in Figure~\ref{fig:studentresponses}(d,e) demonstrates that students valued the work and prefer to do a reproduction more often. Results suggest that having a reproducibility project teaches skills considered important by both student and teacher.

% \begin{figure*}[t]
% \centering
% \includegraphics[width=\linewidth]{other_questions} 
% \caption{Responses to survey questions by  57 students and 87 self-identified AI researchers. The survey question is in the sub-caption. Researchers and students agree that: Reproductions are valuable (a, d, g), that an online repository adds value (b, c), and that more courses should use a reproduction project (h). Researchers differ from students in that researchers more strongly find a reproduction valuable (a), and would consult online reproductions more (d). Researchers think a reproduction is valued less by the community (e) and are less likely to contribute with reproductions (f). Students and researchers both do not agree among themselves if a new paper is more valuable then a reproduction (i), suggesting that the answer is `it depends'. We conclude that the researcher survey respondents welcome an online repository for teaching and structuring reproducibility.}
% \label{fig:allresponses}
% \end{figure*}

\begin{figure}[ht!]
\centering
\includegraphics[width=\linewidth]{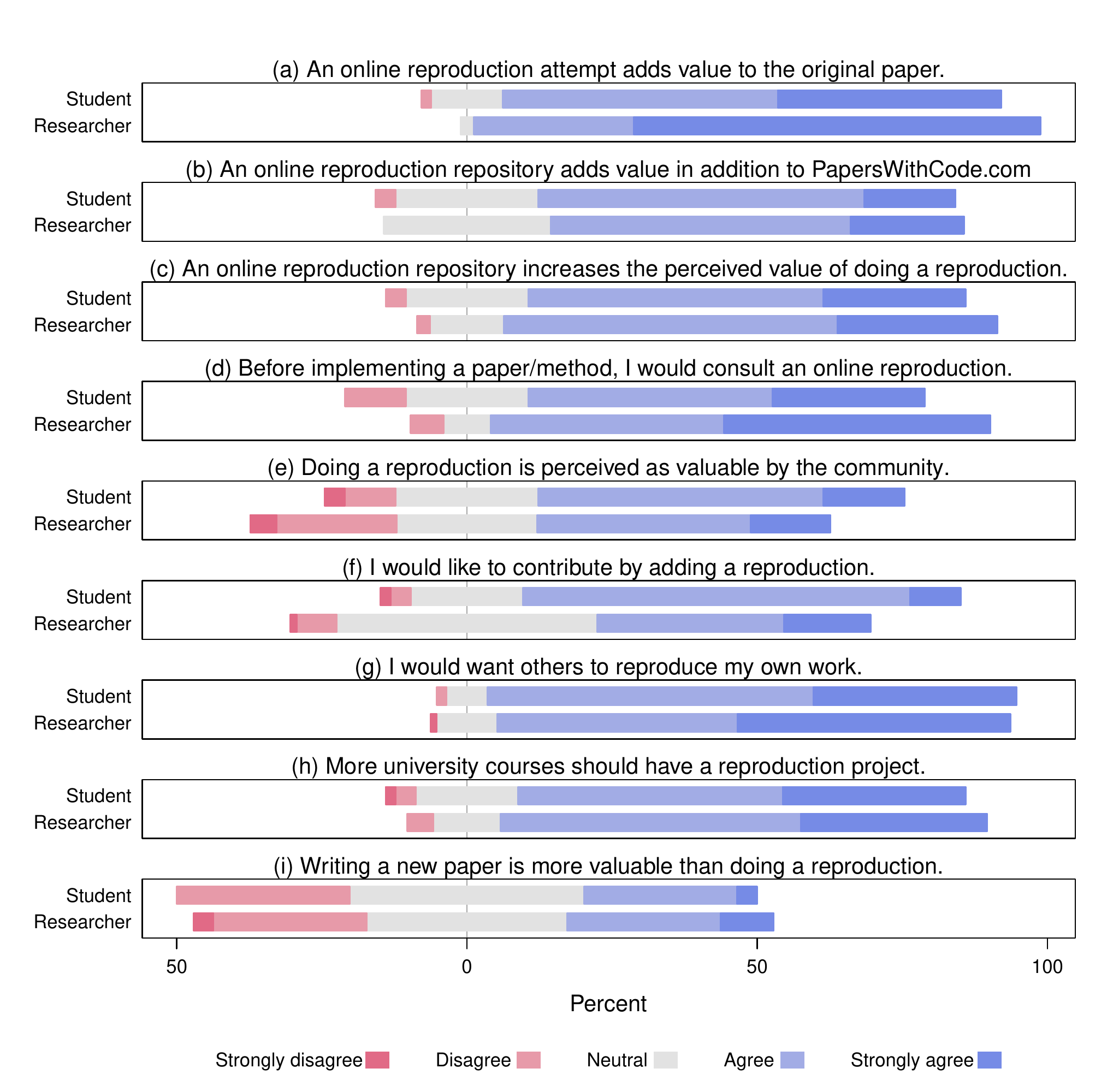} 
\caption{Responses to survey questions by  57 students and 87 self-identified AI researchers. The survey question is in the sub-caption. Researchers and students agree that: Reproductions are valuable (a, d, g), that an online repository adds value (b, c), and that more courses should use a reproduction project (h). Researchers differ from students in that researchers more strongly find a reproduction valuable (a), and would consult online reproductions more (d). Researchers think a reproduction is valued less by the community (e) and are less likely to contribute with reproductions (f). Students and researchers both do not agree among themselves if a new paper is more valuable then a reproduction (i), suggesting that the answer is `it depends'. We conclude that the respondents welcome an online repository for teaching and structuring reproducibility.}
\label{fig:allresponses}
\end{figure}

\subsection{Evaluating the AI researcher survey respondents}

Figure~\ref{fig:allresponses} shows results for the third party AI researchers. We found the following.

\textbf{The AI researcher survey respondents find online reproductions valuable.} 
Results in Figure~\ref{fig:allresponses}(a,d,g) show that students and, especially, researchers find an online reproduction valuable and useful. According to Figure~\ref{fig:allresponses}(i), there is no clear preference for doing a reproduction or writing a paper. Figure~\ref{fig:allresponses}(e) suggests that the perceived value of reproduction by the community is smaller for researchers than for students. 

\textbf{The AI researcher survey respondents find an online reproduction repository valuable.}
Results in Figure~\ref{fig:allresponses}(b,c) show that students and researchers appreciate an online reproduction repository. Figure~\ref{fig:allresponses}(f) shows that researchers are less likely than students to help contribute by doing reproductions.

\textbf{The AI researcher survey respondents see an educational role for courses where students do a reproduction project.}
Results in Figure~\ref{fig:allresponses}(h) show that researchers and students agree that reproduction projects should be used more often in courses.

\smallskip

\noindent Additionally, we make the following observations from Figure~\ref{fig:allresponses}: 

(i) When compared to students, the \emph{researchers think the community values reproductions less (e) and want their own team to work on reproductions less (f)}. This may suggest an inverse relationship between perceived value and willingness to contribute. Yet, when comparing researchers against themselves, most  think the community values reproductions, and most researchers would like to contribute.

(ii) More researchers \emph{want their work reproduced (g) than that they are willing to contribute (f)}. Can we place our hope on the students as future researchers, as they are much more willing to contribute? 

(iii) There is a clear consensus that \emph{reproductions are valuable (a, d, g, i) but some researchers feel that the community does not reward it enough (e)}. Therefore, an important question is how we can change the perception of a low reward for doing reproductions, beyond repositories as reported on here.
\section{Discussion and conclusions}

It should be clear that our results and corresponding analysis are rather preliminary. We are convinced, however, that they warrant low-barrier and long-term solutions accommodating research reproduction. Our \webname~provides one such outlet.  We hope that future analysis of the further accumulated survey data may sketch an even clearer picture. We hope others consider reproducing our effort.

The main conclusions that we draw at present are the following three. 1. Doing a reproduction course project aligns well with learning objectives, and students find it a positive experience.
2. A reproducibility project improves the perceived value of reproductions, and allowing students to blog online about their reproduction project offers an extra incentive to do a reproduction. 3. AI researcher survey respondents are positive about online reproductions and a reproduction repository.

We finally call on the community to add their reproductions to the website \webname~and deploy it in courses: may the next generation of machine learners be reproducers.

%Hayley WIP{I'd like to put a discussion here about experiment validity (missing tests datasets, additional cross validated results ... further sanity checking/triangulation of results.... not sure if this belongs elsewhere. Jan mentioned about minor experiments in the intro but I don't know if those are in a clearly defined category (on the reproduced papers page there's talk of extra experiments but I feel that a strong categorisation of those would really help in moving us forward to more clearly defined evaluation standards in general.. which leads me to the next discussion point about talking about system evaluation and best practices in general - this goes beyond the reproducibility aspect and could become de-emphasised in a course project focusing on reproducbility... or does it...? Something to discuss. :). }

\bibliographystyle{splncs04}
\bibliography{paper}
\end{document}